\documentclass[iop]{emulateapj}
\usepackage{graphicx,natbib}
\usepackage{amsmath,amssymb,latexsym}
\usepackage{multirow}
\usepackage{natbib}
\usepackage{xcolor}
\definecolor{olive}{rgb}{0.5,0.5,0.0}

\newcommand{\bra}{\langle}
\newcommand{\ket}{\rangle}

\shorttitle{Linear Mixing of Planetary Ices}
\shortauthors{Bethkenhagen et al.}

\begin{document}

\title{Planetary Ices and the Linear Mixing Approximation} 
\author{M. Bethkenhagen\altaffilmark{1,2} and E.~R.~Meyer\altaffilmark{3} and S.~Hamel\altaffilmark{2} and N.~Nettelmann\altaffilmark{1,4} and M.~French\altaffilmark{1}
	and L.~Scheibe\altaffilmark{1} and C.~Ticknor\altaffilmark{3} and L.~A.~Collins\altaffilmark{3} and J.~D.~Kress\altaffilmark{3} and J.~J.~Fortney\altaffilmark{4} 
	and R.~Redmer\altaffilmark{1}}

\altaffiltext{1}{Universit{\"a}t Rostock, Institut f{\"u}r Physik, D-18051 Rostock, Germany}
\altaffiltext{2}{Lawrence Livermore National Laboratory, Livermore, California 94550, USA}
\altaffiltext{3}{Theoretical Division, Los Alamos National Laboratory, Los Alamos, New Mexico 87545, USA}
\altaffiltext{4}{Department of Astronomy and Astrophysics, University of California, Santa Cruz, CA 95064, USA}

\begin{abstract}
The validity of the widely used linear mixing approximation for the equations of state (EOS) of planetary ices is investigated at pressure-temperature conditions typical for 
the interior of Uranus and Neptune. The basis of this study are {\it ab initio} data ranging up to 1000~GPa and 20~000~K calculated via density functional theory molecular 
dynamics simulations. In particular, we calculate a new EOS for methane and EOS data for the 1:1 binary mixtures of methane, ammonia, and water, as well as their 2:1:4 
ternary mixture. Additionally, the self-diffusion coefficients in the ternary mixture are calculated along three different Uranus interior profiles and compared to the 
values of the pure compounds. We find that deviations of the linear mixing approximation from the results of the real mixture are generally small; for the thermal EOS they 
amount to 4\% or less. The diffusion coefficients in the mixture agree with those of the pure compounds within 20\% or better. Finally, a new adiabatic model of Uranus with 
an inner layer of almost pure ices is developed. The model is consistent with the gravity field data and results in a rather cold interior ($T_{core}\sim 4000$~K). 
\end{abstract}


\section{Introduction}
\label{sec:Intro}
Our common understanding of the internal composition of the giant planets Uranus and Neptune suggests that the ice-forming volatiles CH$_4$, NH$_3$, and H$_2$O make up a 
major contribution to the total mass of the planets. These planetary ices reside predominantly in the deep interior characterized by hundreds of gigapascal and temperatures 
of several 1000~K. Therefore, many observed properties of these planets, such as luminosity, gravitational moments, and magnetic field, are thought to be determined by the 
physical and chemical properties of these compounds. Hence, the equations of state (EOS) for mixtures of planetary ices under high pressures and temperatures play a crucial 
role for the description of their interior structure.

Since the pioneering work of DeMarcus~\citep{DeMarcus1958} and Peebles~\citep{Peebles1964}, models for giant planets typically employ EOS data for different materials by 
mixing them linearly at constant pressure and temperature. This procedure allows to determine, e.g., the density,
\begin{equation}
\label{LM-rho}
  \frac{1}{\varrho_{LM}(p,T)} = \sum_{i = 1}^N \frac{x_i}{\varrho_i(p,T)}
\end{equation}
and the specific internal energy,
\begin{equation}
  u_{LM}(p,T) = \sum_{i = 1}^N x_i u_i(p,T)
\end{equation}
of a mixture comprised of $N$ components. The density, specific internal energy, and mass fraction of each individual component $i$ are denoted by $\varrho_i$, $u_i$, and 
$x_i$, respectively, in the above equations. 

For the light constituents of giant planets, hydrogen and helium, the validity of the linear mixing approximation (LMA) has been tested in various studies. Early work was based
on Monte Carlo plasma models~\citep{Hubbard1972, Stevenson1975} and was extended to density functional theory studies~\citep{Vorberger2007, Wang2013} later. The later 
studies also investigated several mixing rules for the H-He mixture and found the additive volume rule at constant pressure and temperature to perform best. Recently, \citet{Soubiran2016} 
tested the LMA for the addition of heavy elements to a H-He mixture typical for Jupiter in the dilute limit. Moreover, \citet{Danel2015} investigated the LMA for C$_2$H$_3$ 
as a mixture of carbon and hydrogen for temperatures between 1~eV and 1000~eV finding the strongest deviations of up to 15\% in pressure at 1~eV for small densities. Further
plasmas have been studied at such high temperatures regarding various mixing rules \citep{Lambert2008, Horner2008, Magyar2013}. 

Phase transitions might cause strong deviations of the linear mixing model from the behavior of the real mixture. For instance, water is predicted to become superionic along 
the adiabat of Uranus and Neptune~\citep{Cavazzoni1999,French2009,Redmer2011}, whereas ammonia remains fluid~\citep{Cavazzoni1999, Bethkenhagen2013}. There is also 
experimental indication for superionic behavior in water~\citep{Goncharov2005,Sugimura2012} and ammonia~\citep{Ninet2012}, albeit these experiments were made at temperatures 
far below the isentropes of Uranus or Neptune. Methane, however, does not become superionic but is instead predicted to form long-chained hydrocarbons or potentially to 
demix into carbon and hydrogen under the conditions present in Uranus~\citep{Hirai2009, Gao2010, Spanu2011, Lobanov2013}. One can thus suspect strong deviations to occur 
between real and linear (ideal) mixing behavior, at least in certain regions of the pressure-temperature space.

Selected real icy mixtures have been investigated previously using {\it ab initio} simulations, e.g., various H$_2$O-NH$_3$ compositions~\citep{Bethkenhagen2015, Jiang2017, Naden-Robinson2017}, 
H-H$_2$O~\citep{Soubiran2015}, H-He with heavy element enrichment~\citep{Soubiran2016}, and the H-C-N-O mixture~\citep{Chau2011}. There also exist 
experimental data (EOS and electrical conductivity) for H-C-N-O mixtures, most of them based on shock compression experiments \citep{Radousky1990, Nellis1997,Chau2011}, but 
their conditions are limited to single compression paths and pressures less than 200~GPa. In general, EOS data for icy mixtures at extreme pressures and temperature are 
still sparse, and the validity of the LMA for molecular compounds has not yet been systematically checked. It is the purpose of this work to provide such a systematic study 
across a wide range of pressure-temperature conditions. 

This paper is organized as follows. In the first part of Section \ref{sec:EOS} we briefly describe the interior profiles of Uranus which serve as a guide for the 
thermodynamic conditions of interest here. The second part of Section \ref{sec:EOS} describes the computational method used and discusses the EOS of the molecular compounds 
and their mixtures. Section 3 investigates the performance of the LMA for the binary mixtures and a ternary mixture at representative Uranus interior conditions. 
Section \ref{sec:Diffusion} investigates the self-diffusion coefficients in the mixture. In Section \ref{sec:IcyUranusModel} we apply our EOS for the planetary ices to Uranus.
A summary and final conclusions are given in Section 6.  

\section{Equations of State (EOS)}
\label{sec:EOS}

\subsection{Ice giant interior profiles}\label{sec:Uprofiles}

We compute the equations of state of ices and their mixtures at pressure-temperature conditions relevant for the interior of Uranus and Neptune. However, these planets' 
internal temperature profiles cannot be measured and thus are not well known. Most common models assume an adiabatic interior based on the idea that at least the part 
generating the magnetic field should be convective \citep{Soderlund2013} and thus nearly adiabatic. Those models suggest a range from cold $(T_{\rm core}\sim 2000$~K) to warm 
$(T_{\rm core}\sim 6000$~K) interiors depending on the chosen underlying materials and consequently equations of state \citep{HubbardMacfarlane1980,Redmer2011, Nettelmann2013}.
Cold interiors may result from a cold-start formation, in particular for the case of Uranus \citep{Hubbard1995}. Likewise, a high ice content might also originate from 
the planetary formation process \citep{PodRey1987}. On the contrary, hot interiors with temperatures exceeding 10~000~K are obtained from models including a strongly 
super-adiabatic region \citep{Nettelmann2016}. Such a thermal boundary layer can occur if the barrier between H/He-dominated atmosphere and icy interior inhibits the heat 
flow across it \citep{Nettelmann2016}.

Due to this uncertainty, we consider here three interior profiles from different Uranus models:

$(i)$ the warm ($T_{\rm core}\sim 6000$~K) adiabatic Uranus model of \citet{Redmer2011}, where ices are represented solely by a water EOS and the resulting size of the ionic 
water region is found to be consistent with predictions from magnetic field models,

$(ii)$ the hot ($T_{\rm core}\sim 14~000$~K) class III Uranus model with thermal boundary layer (TBL) by \citet{Nettelmann2016}, which can explain the current faintness of 
Uranus by equilibrium evolution with the solar incident flux, 

($iii$) a rather cool ($T_{\rm core}\sim 4000$~K) icy model which assumes an adiabatic interior of a mixture of methane, ammonia, and water with only a tiny fraction (1\%) 
of hydrogen and helium needed to explain the gravity field, for details see Section \ref{sec:IcyUranusModel}).

Hereafter, these models are respectively labeled \textit{water-only, TBL,} and \textit{icy}. They serve as representative guide for our EOS calculations that aim to cover 
typical pressure-temperature conditions inside Uranus.

\subsection{Density Functional Theory Molecular Dynamics Simulations (DFT-MD)}\label{sec:DFT}

The entire set of EOS data was obtained with the Vienna {\it Ab Initio} Simulation Package (VASP)~\citep{Kresse1993a, Kresse1993b, Kresse1996, Hafner2008}. This 
DFT-MD code is based on the Born-Oppenheimer approximation and describes the electron system via density functional 
theory (DFT) at finite temperatures~\citep{Hohenberg1964, Kohn1965, Mermin1965, Weinert1992, Wentzcovitch1992}, while the ions are propagated as classical particles within 
a molecular dynamics (MD) framework. We control the ionic temperature within the NVT ensemble by employing a Nos\'e-Hoover thermostat~\citep{Nose1984,Hoover1985}. 
The interaction between electrons and ions is described by projector augmented wave (PAW) pseudopotentials~\citep{Blochl1994, Kresse1999}. The approximation of Perdew, 
Burke, and Ernzerhof (PBE) was chosen for the exchange-correlation functional~\citep{Perdew1996}.

In general, the following parameters have been used throughout all simulations, if not stated otherwise. The plane-wave energy cutoff was set to 1000~eV and the Baldereschi 
mean-value point was used to sample the $\vec{k}$ space in most simulations. Particle numbers varied between 16 and 84 molecules, depending on the composition, density, and 
temperature. The simulation duration was typically between 10~ps and 20~ps after equilibration with timesteps between 0.25 and 0.4~fs. All simulation parameters have been 
thoroughly checked to ensure the convergence of our results.

In the following, we describe our present EOS database, which contains, to a great extent, novel data that are consistently complemented with values previously reported in 
literature. The resulting EOS data for pressure $p(\varrho,T)$ and internal energy $u(\varrho,T)$ cover a grid up to temperatures of 20~000~K and pressures of 1000~GPa.

The specific internal energies of the pure compounds were shifted to zero at a reference point of 1000~GPa and 20~000~K. The specific internal energies of the binary and 
ternary mixtures were then renormalized with the same shifts applied to the pure compounds weighted by their respective mass fractions. This renormalization aids in the 
visual interpretation of the data but does not change the physics.

\subsection{Methane}\label{sec:methane}

Although there exist various EOS of methane (e.g., \citealp{Kerley1980, Setzmann1991, Sherman2012}), none of them covers the entire pressure-temperature region required for 
Uranus and Neptune interior models. We have therefore computed a new methane EOS using DFT-MD simulations. We simulated 54 molecules in the simulation box, which were 
initially placed on a bcc lattice for every simulation run to avoid a bias toward certain molecular configurations, such as polymers.

\begin{figure}[htb]
  \vspace{-4pt}
  \includegraphics[width=1.0\columnwidth]{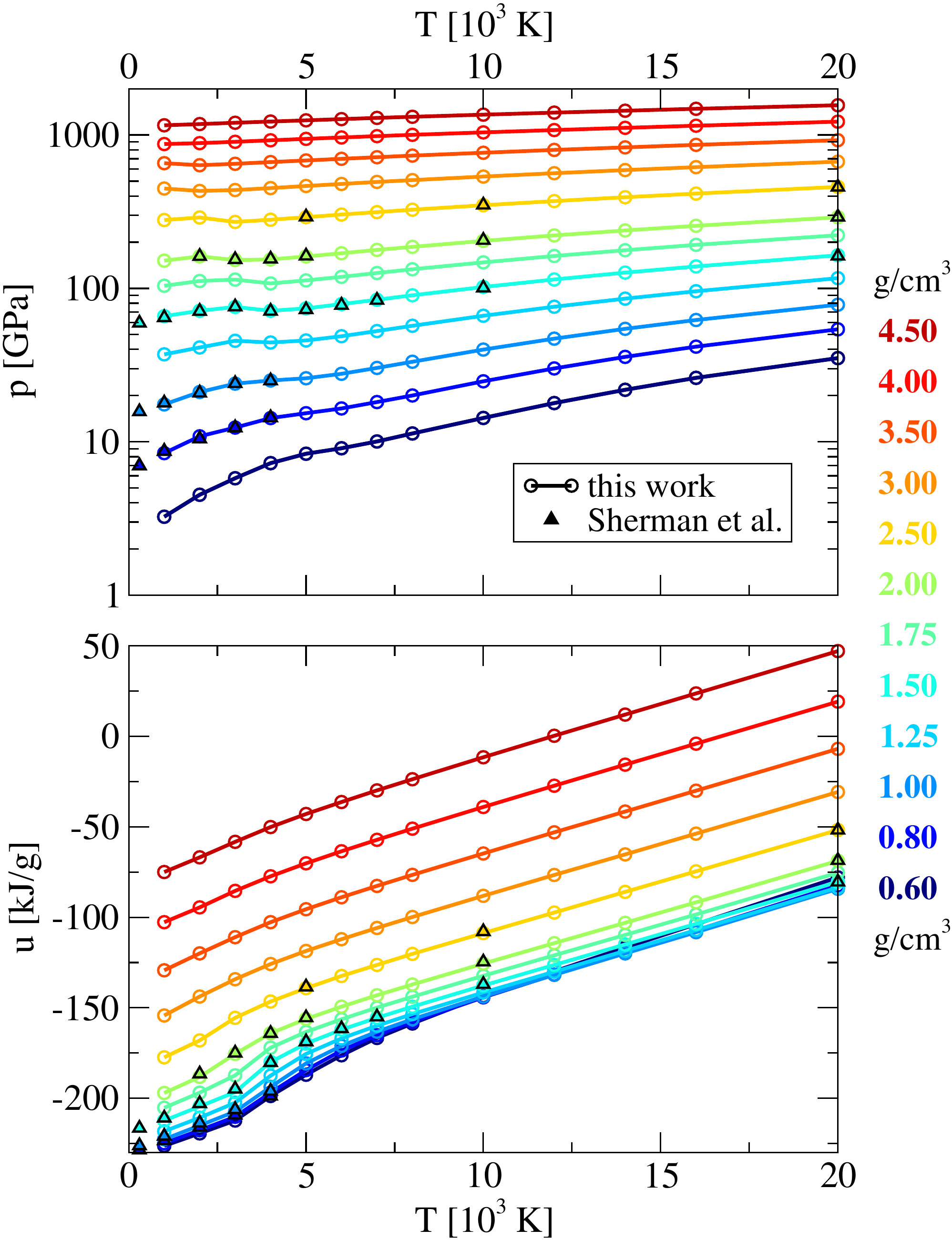}
  \caption{Thermal (top panel) and caloric (bottom panel) EOS of methane shown along isochors (solid lines with circles) in comparison to the EOS data by~\citet{Sherman2012} 
  (colored triangles). Every circle represents a DFT-MD run of at least 10~ps.}
  \label{fig:methane}
\end{figure}

The resulting EOS is shown along isochores in Fig.~\ref{fig:methane} in comparison with earlier work by~\citet{Sherman2012}, which is based on the same {\it ab initio} 
simulation technique as applied here with slightly different simulation parameters. Most of our data agree within 2\% in pressure and within 1~kJ/g in specific internal 
energy compared to the results of~\citet{Sherman2012}. However, we find more significant deviations of up to 4\% and up to 2~kJ/g below 4000~K associated with the melting 
and dissociation of CH$_4$ indicated by the change in slope in in Fig.~\ref{fig:methane}. Performing additional heating and cooling simulations in this region, we observed 
the formation of different molecular species. While small molecules like ethane can be identified at small densities, we find polymers starting to form at higher densities   
and temperatures above 3000~K. These effects will be addressed in more detail in future work on the phase diagram of methane with special emphasis on potential demixing 
into carbon and hydrogen. The present methane EOS contains these effects via the standard procedure of time-averaging simulation data.

\subsection{Ammonia}

We extended the ammonia EOS data set from~\citet{Bethkenhagen2013}, which ranged up to 10~000~K and 330~GPa, to higher pressures and temperatures. To ensure full consistency 
with all other simulation data from this work, the correction due to nuclear quantum effects was removed from the published data set \citep{Bethkenhagen2013}. In particular, 
the density grid was extended by seven additional densities per isotherm in order to cover the pressure range up to 1000~GPa. Additionally, four more isotherms (12~000~K, 
14~000~K, 16~000~K, and 20~000~K) were computed.

\subsection{Water}

The water EOS is based on simulations used to generate the tabular data set of~\citet{French2009}, which were extended in runtime to at least 10~000 timesteps to reduce 
their statistical uncertainty. The data set was extended to lower densities (0.2~g/cm$^3$) using simulations with 16, 24, or 54 molecules and the $\Gamma$ point. Simulations 
at densities of 4~g/cm$^3$ and higher were rerun with the Baldereschi mean-value point to further improve the numerical convergence. Finally, the region of the superionic 
phase was filled with the raw data from more recent simulations for superionic water with a bcc oxygen lattice~\citep{French2016}.

\subsection{1:1 Binary Mixtures}\label{sec:RM_bin}

We have calculated the EOS of the two 1:1 binary mixtures water-methane and ammonia-methane and extended the 1:1 water-ammonia data of~\citep{Bethkenhagen2015} 
to higher temperatures and lower densities. The latter mixture has been calculated using 32 molecules and the Monkhorst-Pack $2\times 2\times 2$ grid, while the former two 
mixtures have been calculated with 54 molecules and were started with a molecular bcc lattice. The simulations for all three binary mixtures have been carried out on the 
same temperature grid as for the pure compounds. In total, 13 temperatures were considered (1000~K -- 8000~K: 1000~K steps, 10~000~K -- 16~000~K: 2000~K steps, 20~000~K) 
with each isotherm containing at least ten density points.

\begin{figure}[htb]
   \vspace{-4pt}
   \includegraphics[width=1.0\columnwidth]{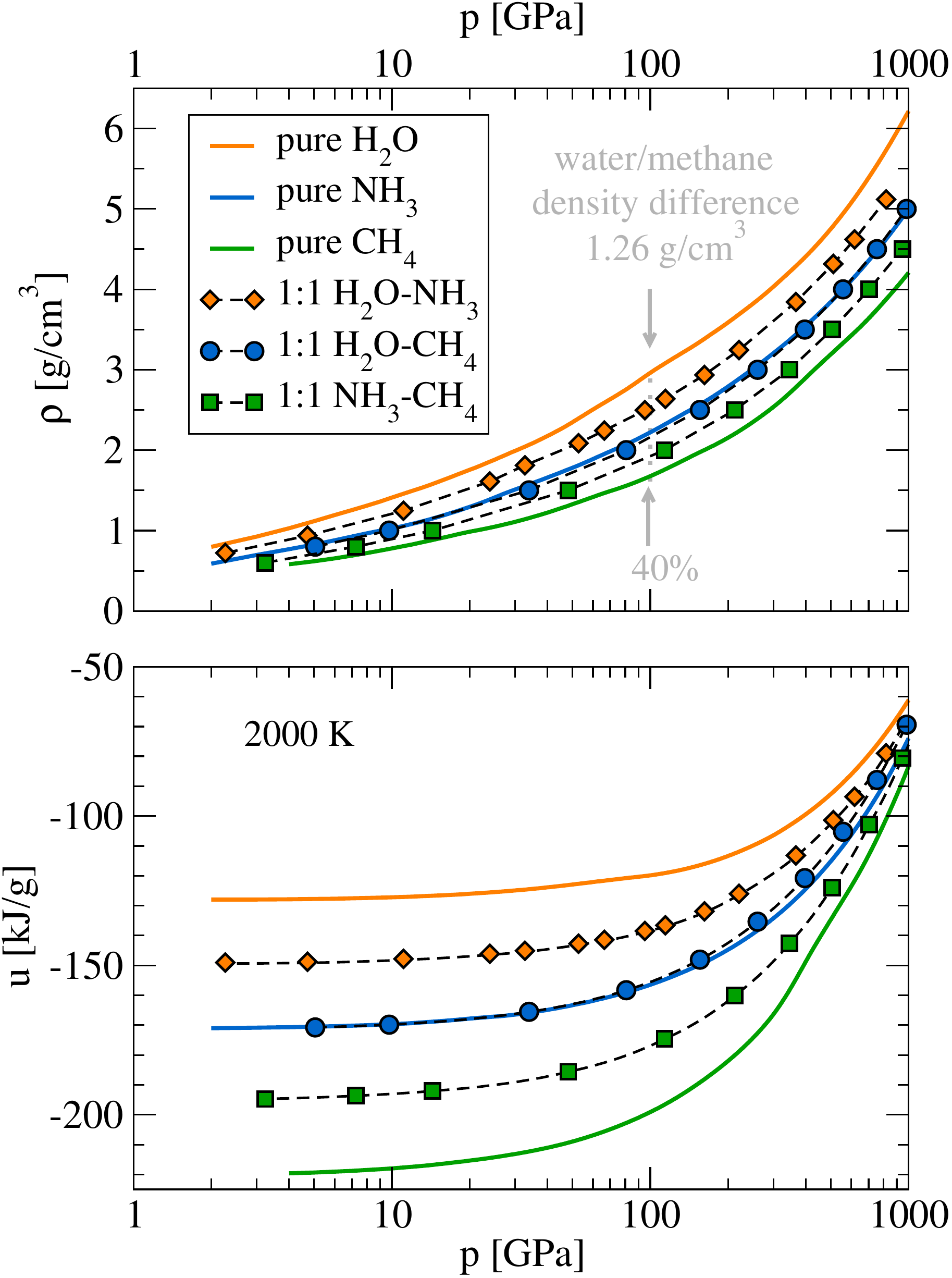}
   \caption{Density (top panel) and internal energy (bottom panel) vs. pressure for the pure compounds (solid lines) in comparison to results from the real 1:1 binary 
    mixtures (dashed lines with filled symbols) along the representative 2000~K isotherm. 
   \label{fig:binary-2000K}}
\end{figure}

The 2000~K isotherm is shown in Fig.~\ref{fig:binary-2000K} as an example for each of the 1:1 binary mixtures as well as for the pure compounds. The considered materials 
behave systematically with mean molecular weight ${m}$, so that an increase in $m$ directly translates into a proportional density increment at a given pressure. For example,
the difference between the lightest material, CH$_4$, and the heaviest material, H$_2$O, amounts to 1.26~g/cm$^3$ at 100 GPa and 2000~K, i.e., the density difference is 
about 40\%. Hence, the composition of the icy mixture has an appreciable impact on its resulting thermodynamic properties. Furthermore, the densities of pure ammonia 
($m=17$ g/mol) and of the 1:1 water-methane mixture ($m=17$ g/mol), agree within 3\% along the 2000~K isotherm, emphasizing the systematic behavior with mean molecular 
weight of the thermal EOS.

A very similar picture is obtained for the specific internal energy, see lower panel of Fig.~\ref{fig:binary-2000K}. This suggests that the energetics of the mixtures are 
mainly determined by the average number of bonds (or degree of association and correlation) between hydrogen and heavy nuclei. Dissociation of the molecules occurs across 
the same pressure and temperature ranges for the pure substances as well as in their mixtures~\citep{Meyer2015}.  

\subsection{Ternary 2:1:4 Methane-Ammonia-Water Mixture}\label{sec:RM_ter}

Prior to this work an extensive study has been performed on different concentrations of the ternary mixtures in order to gain an insight on the structural and chemical
properties of those mixtures~\citep{Meyer2015}. We chose the 2:1:4 mixture for this study since it provides a good compromise between computational effort and resemblance 
to the solar abundances of about 4:1:7 of C:N:O \citep{Asplund2009}. We simulated mixtures containing 24 methane, 12 ammonia, and 48 water molecules. This was performed 
along three planetary $P$--$T$ profiles of Uranus (see Section \ref{sec:Uprofiles}). For temperatures below or equal to 6000~K the reciprocal space was sampled with the 
Baldereschi mean-value point, while for higher temperatures the Monkhorst-Pack $2\times 2\times 2$ grid was used. Each simulation run was started from a density as derived 
using the EOS of the pure compounds and the LMA. Every 1000 timesteps the pressure was checked and the volume of the simulation box adapted until the desired pressure was 
matched up to a deviation of 2\%. Since this procedure is computationally expensive, especially for low pressures, we typically chose two different volumes and interpolated 
linearly between the results in order to match target pressures below 40~GPa.


\section{Testing the Linear Mixing Approximation for Density and Energy}\label{sec:LMA}

\subsection{Binary Mixtures}\label{sec:Binary}

We define the deviations in density and specific internal energy between the linear (LM) and the real mixtures (real) through the quantities:
\begin{align}
  \Delta \varrho(p,T) &= \varrho_{LM}(p,T)/ \varrho_{real}(p,T) - 1\text{ ,}
\end{align}
and
\begin{align}
  \Delta u(p,T) &= u_{LM}(p,T) - u_{real}(p,T) \text{ ,}
\end{align}
respectively. 

\begin{figure*}[!t]
  \vspace{-4pt}
  \includegraphics[width=1.0\textwidth]{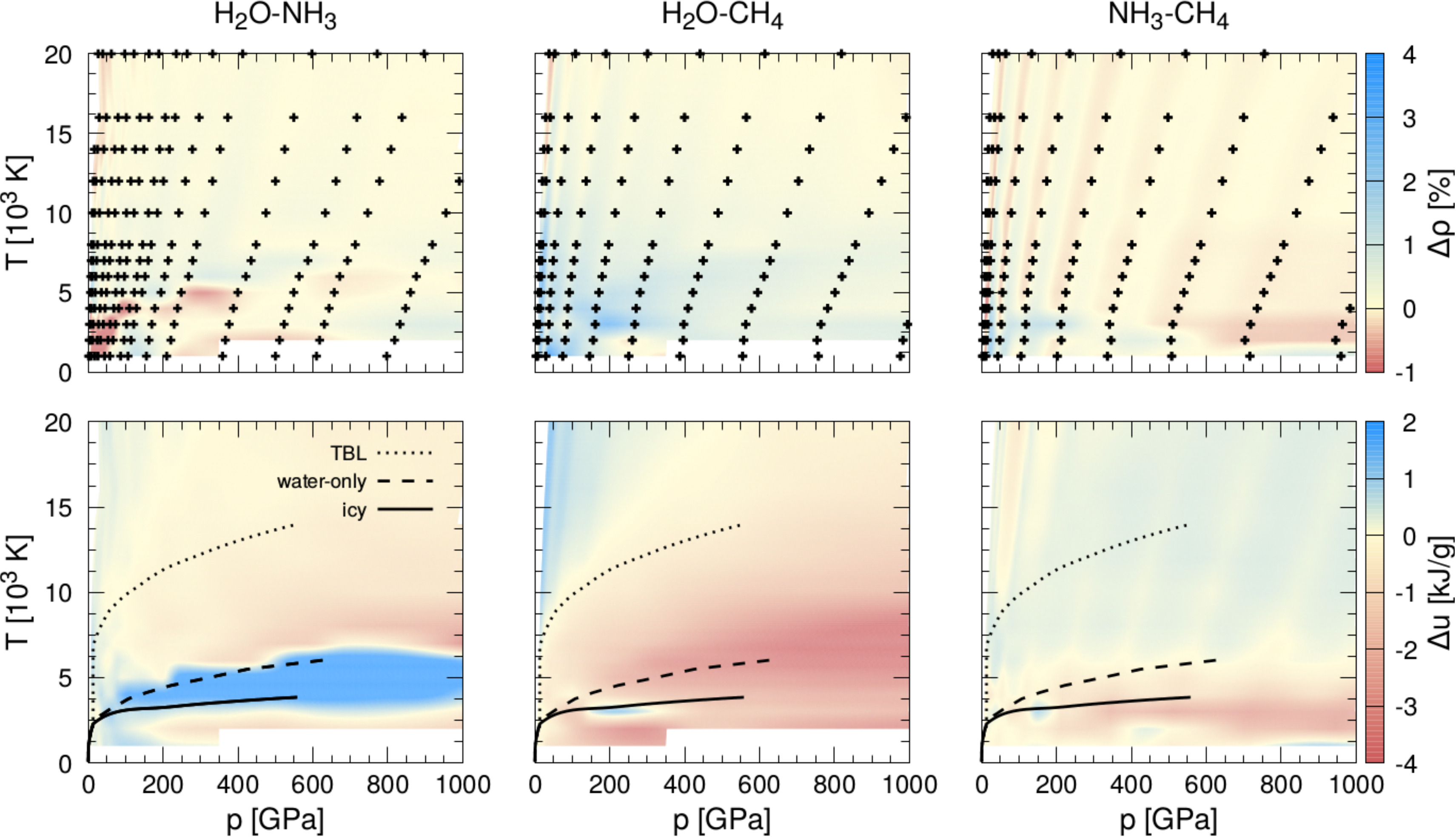}
  \caption{Deviation of densities (upper panels) and internal energies (lower panels) of the considered real binary 1:1 mixtures from the linear mixing approximation. 
	   The color code is the same for the three mixtures H$_2$O-NH$_3$ (left), H$_2$O-CH$_4$ (middle), and NH$_3$-CH$_4$ (right). The underlying data for 
	   the real binary mixtures are shown as black crosses in the upper panels. For reference, the temperature profiles of the {\it TBL}~\citep{Nettelmann2016}, 
	   {\it water-only}~\citep{Redmer2011}, and {\it icy}~(see Sec. \ref{sec:IcyUranusModel}) Uranus models (black dotted, dashed, and solid lines respectively) 
	   are displayed as well in the lower panels.
	  \label{fig:binary-all}}
\end{figure*} 

All EOS data for the pure compounds and for the binary mixtures were interpolated onto a common rectangular pressure-temperature grid. In particular, the pressures along 
isotherms were interpolated using Akima splines, while the temperatures were interpolated linearly along isobars. 

Figure \ref{fig:binary-all} shows the results of this comparison. The deviation in density varies between -1\% and 4\%, while that of the specific internal energy amounts to 
values between -4~kJ/g and 2~kJ/g. Most important, the largest deviations in internal energy occur when at least one but not all of the pure compounds become superionic (here
water and ammonia) so that its heavy particles form a crystal which releases lattice formation energy. In case of the water-ammonia mixture, we find a slightly negative 
deviation in ∆u (red stripe above blue region in Fig. 3) as the superionic phase of water forms below 10 000 K (French et al. 2009). This behavior is inverted (blue region) 
when an oxygen-nitrogen lattice appears also in the 1:1 mixture at few 1000 K colder temperatures (Bethkenhagen et al. 2015). In addition to pure water and the 1:1 
water-ammonia mixture, pure ammonia becomes superi onic as well but only below 4000 K (Bethkenhagen et al. 2013), so that the LMA is fulfilled well again in the cool, dense 
mixture (disappearance of blue region). The effect of superionicity on the density deviations is much weaker because the density jumps to the fluid phase is relatively small.
Note that although the lattice type is different in water (bcc), ammonia (fcc), and in the mixture (P4/nmm, Ima2, Pma2, Pm), this does not lead to any specific effects on 
$\Delta u$ or $\Delta \varrho$.

In case of the binary mixtures containing methane, the occurrence of superionic phases in water or ammonia leads to similar deviations in $\Delta u$ as discussed above. Pure 
methane does not become superionic but instead decomposes into long-chained molecules in our simulations. Similar molecular aggregates can occur also in mixtures with water 
and ammonia~\citep{Meyer2015}. The change in internal energy due to these chemical reactions has the same sign as the formation of lattices in the water-ammonia system, so 
that these two very different phenomena lead to a partial compensation in $\Delta u$ at low temperatures. 

For the density deviations in the methane-containing mixtures the picture is less clear. The slightly visible maxima and minima in $\Delta \varrho$ cannot be 
directly related to specific phase transitions, but rather might be the result from our interpolation using Akima splines. These third order polynomials can lead to an 
oscillatory behavior when calculating small differences, especially when the underlying data grid is coarse and/or the data possess residual statistical fluctuations. This 
is supported by the fact that our EOS database for water-ammonia is larger compared to the other two mixtures for pressures below 200~GPa as indicated by 
the black crosses in the upper panels in Fig.~\ref{fig:binary-all}. Hence, the sparser datasets for the water-methane and ammonia-methane mixtures lead to more pronounced 
oscillations. Therefore, we see the density deviation of up to 4\% as observed here as upper limit to the true performance of the LMA.

Nevertheless, we can draw two clear conclusions from our systematic study:

First, {\it the LMA for density and internal energy works very well} for all three binary mixtures {\it if the same thermodynamic phase is present} both in the pure compounds 
and in the real mixture. The numerical deviations are often less than one percent and never larger than 4\%. This applies to the fluid as well as to the superionic phases. In 
the latter case, it is even of little importance which type of lattice is present, which was also observed in an earlier study on pure water~\citep{French2016}.  

Second, deviations from the LMA can be attributed to the formation of nitrogen-oxygen lattices or to prominent chemical reactions involving carbon. Their quantiative effect 
amounts to -1\% and 4\% for density and -4~kJ/g and 2~kJ/g for the internal energy. The latter corresponds to a maximum change of only 200~K or 4\% in radius along a typical
Uranus isentrope.   

\subsection{Ternary Mixture}\label{sec:Ternary}

\begin{figure}[hbt]
  \vspace{-4pt}
  \includegraphics[width=1.0\columnwidth]{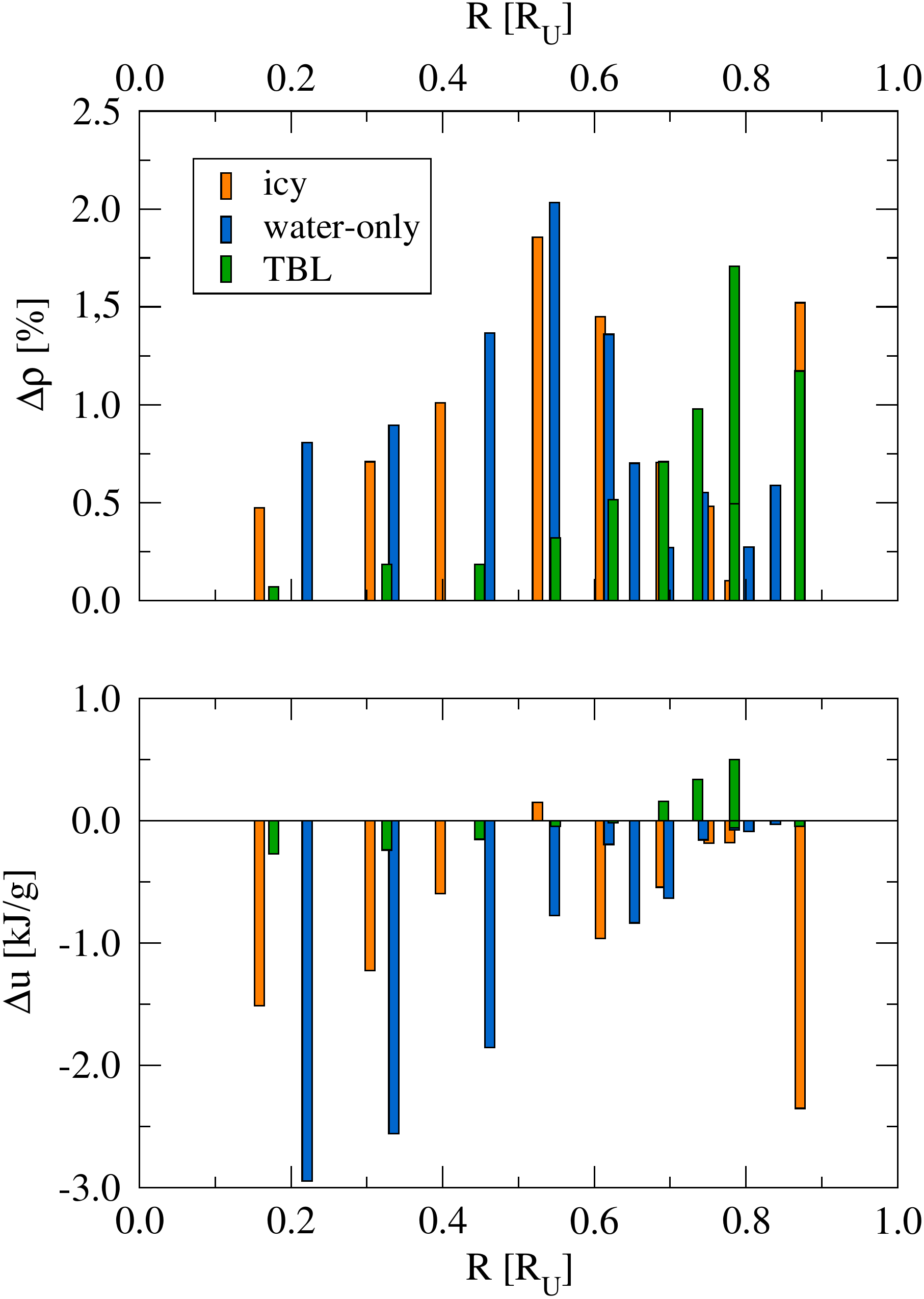}
  \caption{Deviations of density (upper panel) and specific internal energy (lower panel) between the real 2:1:4 methane-ammonia-water mixture (solid lines) and the LMA 
  along the three planetary profiles of Uranus.} 
   \label{fig:ternary}
\end{figure}

The results for the real 2:1:4 ternary mixture along the three considered planetary profiles are shown in Fig.~\ref{fig:ternary} and compared with the LMA. 

In contrast to the binary mixtures, the LMA tends to systematically overestimate densities and, at the same time, underestimate the internal energy in the ternary 
mixture. However, the magnitude of the deviations is small, and we find the LMA to perform even slightly better for the ternary than for the binary mixtures, which might 
partially result from the simulation procedure used here. The direct simulation of the real mixture along the planetary profiles allows us to avoid additional interpolation 
on a rather coarse grid, which was necessary to investigate the binary mixtures on the entire pressure--temperature plane. 

The maximum deviations in density amount to up to 2.1\% and are very similar for all three Uranus profiles up to 10~000~K. The density deviation is even less than 0.5\%
above this temperature, which is only reached by the hottest profile (TBL). This planetary profile is also clearly standing out in terms of the internal energy deviation due 
to a remarkably small deviation from the linear mixing model of only up to $\pm$0.6~kJ/g. The largest energy deviation of up to -2.9~kJ/g is found for the significantly 
colder adiabatic models. As discussed in Sec.~\ref{sec:Binary}, the more pronounced deviations in the internal energy for the colder models most likely result 
from the formation of superionic phases in water and ammonia as well as from changes in the chemical bonding of carbon atoms. This occurs both in pure as well as in the 
ternary mixture, which has been extensively studied by~\citet{Meyer2015}.

\section{Diffusion Coefficients}\label{sec:Diffusion}

\begin{deluxetable*}{cccccccc}[htb]
  \tablecolumns{8}
  \tablewidth{0pc}
\tablecaption{Self-diffusion coefficients of the real ternary mixture along the considered planetary profiles \label{tab:diff}}
\tablehead{
\colhead{R [R$_{\mathrm{U}}$]} & \colhead{$\varrho$ [g/cm$^3$]} & \colhead{T [K]} & \colhead{p [GPa]} & \colhead{D$_\mathrm{H}$ [cm$^2$/s]} 
& \colhead{D$_\mathrm{C}$ [cm$^2$/s]} & \colhead{D$_\mathrm{N}$ [cm$^2$/s]} & \colhead{D$_\mathrm{O}$ [cm$^2$/s]}} 
\startdata
\cutinhead{\it{icy} planetary profile}
0.872	& 0.881	& 1500	& 4.3	& 3.06$\times10^{-4}$	& 2.58$\times10^{-4}$	& 3.16$\times10^{-4}$	& 3.48$\times10^{-4}$ \\  
0.779	& 1.19	& 2265	& 13.0	& 2.65$\times10^{-4}$	& 2.16$\times10^{-4}$	& 2.47$\times10^{-4}$	& 2.74$\times10^{-4}$ \\
0.751	& 1.45	& 2500	& 24.9	& 2.42$\times10^{-4}$	& 1.58$\times10^{-4}$	& 1.78$\times10^{-4}$	& 1.89$\times10^{-4}$ \\  
0.688	& 1.95	& 2900	& 62.4	& 3.02$\times10^{-4}$	& 7.44$\times10^{-5}$	& 7.97$\times10^{-5}$	& 9.20$\times10^{-5}$ \\  
0.608	& 2.46	& 3150	& 124	& 4.95$\times10^{-4}$	& 3.57$\times10^{-5}$	& 4.51$\times10^{-5}$	& 5.13$\times10^{-5}$ \\  
0.525	& 2.92	& 3250	& 203	& 6.41$\times10^{-4}$	& 2.82$\times10^{-5}$	& 3.06$\times10^{-5}$	& 2.93$\times10^{-5}$ \\  
0.397	& 3.45	& 3500	& 328	& 8.28$\times10^{-4}$	& 2.79$\times10^{-5}$	& 2.31$\times10^{-5}$	& 2.16$\times10^{-5}$ \\  
0.304	& 3.77	& 3650	& 418	& 8.95$\times10^{-4}$	& 2.85$\times10^{-5}$	& 2.17$\times10^{-5}$	& 2.03$\times10^{-5}$ \\  
0.158	& 4.19	& 3850	& 558	& 9.52$\times10^{-4}$	& 3.37$\times10^{-5}$	& 2.05$\times10^{-5}$	& 1.78$\times10^{-5}$ \\
\cutinhead{\it{water-only} planetary profile} 
0.839	& 1.00	& 1775	& 6.9	& 2.83$\times10^{-4}$	& 2.43$\times10^{-4}$	& 2.80$\times10^{-4}$	& 3.05$\times10^{-4}$ \\  
0.804	& 1.11	& 2050	& 10.0	& 2.73$\times10^{-4}$	& 2.30$\times10^{-4}$	& 2.78$\times10^{-4}$	& 2.87$\times10^{-4}$ \\
0.785	& 1.27	& 2375	& 16.0	& 2.45$\times10^{-4}$	& 1.91$\times10^{-4}$	& 2.05$\times10^{-4}$	& 2.48$\times10^{-4}$ \\  
0.744	& 1.56	& 2775	& 31.5	& 2.78$\times10^{-4}$	& 1.51$\times10^{-4}$	& 1.68$\times10^{-4}$	& 1.88$\times10^{-4}$ \\
0.698	& 1.83	& 3125	& 52.7	& 3.96$\times10^{-4}$	& 1.20$\times10^{-4}$	& 1.26$\times10^{-4}$	& 1.53$\times10^{-4}$ \\  
0.653	& 2.10	& 3425	& 79.8	& 5.92$\times10^{-4}$	& 1.00$\times10^{-4}$	& 1.07$\times10^{-4}$	& 1.40$\times10^{-4}$ \\  
0.619	& 2.31	& 3775	& 107	& 8.72$\times10^{-4}$	& 1.07$\times10^{-4}$	& 1.18$\times10^{-4}$	& 1.35$\times10^{-4}$ \\  
0.548	& 2.60	& 4150	& 153	& 1.22$\times10^{-3}$	& 1.19$\times10^{-4}$	& 1.22$\times10^{-4}$	& 1.28$\times10^{-4}$ \\  
0.462	& 3.03	& 4650	& 240	& 1.61$\times10^{-3}$	& 1.39$\times10^{-4}$	& 1.27$\times10^{-4}$	& 1.24$\times10^{-4}$ \\  
0.335	& 3.54	& 5250	& 375	& 1.84$\times10^{-3}$	& 1.89$\times10^{-4}$	& 1.39$\times10^{-4}$	& 1.32$\times10^{-4}$ \\ 
0.221	& 3.96	& 5750	& 510	& 1.94$\times10^{-3}$	& 2.10$\times10^{-4}$	& 1.44$\times10^{-4}$	& 1.31$\times10^{-4}$ \\
\cutinhead{\it{TBL} planetary profile} 
0.871	& 0.911	& 1500	& 4.8	& 2.73$\times10^{-4}$	& 2.42$\times10^{-4}$	& 2.46$\times10^{-4}$	& 3.04$\times10^{-4}$ \\  
0.785	& 1.20	& 2175	& 13.0	& 3.27$\times10^{-4}$	& 2.77$\times10^{-4}$	& 3.12$\times10^{-4}$	& 3.35$\times10^{-4}$ \\
0.785	& 0.91	& 6875	& 13.0	& 5.12$\times10^{-3}$	& 1.62$\times10^{-3}$	& 1.80$\times10^{-3}$	& 1.89$\times10^{-3}$ \\  
0.737	& 1.33	& 8000	& 32.8	& 6.39$\times10^{-3}$	& 1.62$\times10^{-3}$	& 1.63$\times10^{-3}$	& 1.66$\times10^{-3}$ \\  
0.691	& 1.63	& 9000	& 59.0	& 7.31$\times10^{-3}$	& 1.64$\times10^{-3}$	& 1.52$\times10^{-3}$	& 1.48$\times10^{-3}$ \\  
0.625	& 2.04	& 10000	& 110	& 7.63$\times10^{-3}$	& 1.75$\times10^{-3}$	& 1.47$\times10^{-3}$	& 1.35$\times10^{-3}$ \\  
0.549	& 2.42	& 11000	& 176	& 7.19$\times10^{-3}$	& 1.68$\times10^{-3}$	& 1.43$\times10^{-3}$	& 1.23$\times10^{-3}$ \\  
0.449	& 2.85	& 12000	& 273	& 6.88$\times10^{-3}$	& 1.70$\times10^{-3}$	& 1.27$\times10^{-3}$	& 1.19$\times10^{-3}$ \\  
0.326	& 3.30	& 13000	& 398	& 6.30$\times10^{-3}$	& 1.62$\times10^{-3}$	& 1.38$\times10^{-3}$	& 1.17$\times10^{-3}$ \\
0.177	& 3.77	& 14000	& 559	& 5.79$\times10^{-3}$	& 1.49$\times10^{-3}$	& 1.15$\times10^{-3}$	& 1.05$\times10^{-3}$ 
\enddata
\tablenotetext{}{}
\end{deluxetable*}

The self-diffusion coefficients $D_{\alpha}$ were calculated using velocity autocorrelation functions via the expression

\begin{equation}\label{e:diff}
  D_{\alpha} = \lim_{t\rightarrow\infty} \frac{1}{3 N_{\alpha}}\int_0^{t} \sum_{i=1}^{N_\alpha} \bra \vec{v}_{i}(0)\cdot \vec{v}_{i}(\tau)\ket d\tau \text{,}
\end{equation}
where $N_\alpha$ denotes the particle number of species $\alpha$. Here, we compare the self-diffusion coefficients for the species hydrogen, carbon, nitrogen, and oxygen in 
the ternary mixture with those in the pure compounds CH$_4$, NH$_3$, H$_2$O at the same pressure and temperature. An explicit mixing rule for the $D_{\alpha}$, as has been 
suggested for higher temperatures~\citep{Horner2009}, is not examined here.

\begin{figure}[htb]
  \vspace{-4pt}
  \includegraphics[width=1.05\columnwidth]{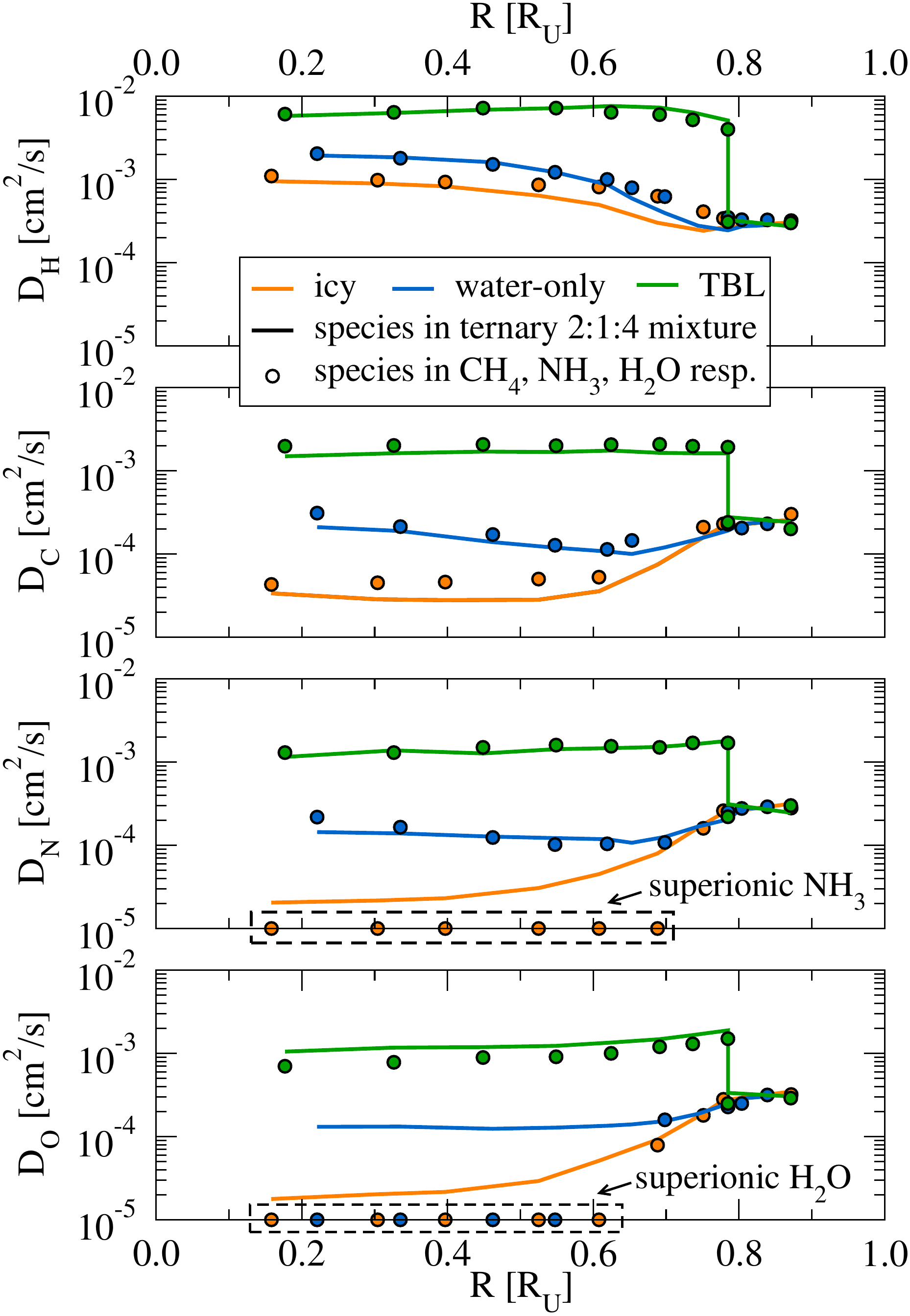}
  \caption{The self-diffusion coefficients of the constituents H, C, N, and O in the real ternary mixture (solid lines) along the planetary profiles as a function of Uranus' 
  radius. Diffusion coefficients framed by dashed boxes are artificial and mark the superionic regime, where the respective 
  diffusion coefficients vanish. The colored circles indicate the diffusion coefficients in the pure compounds shown here exemplarily for three Uranus profiles.
  \label{fig:diff}}
\end{figure}

We computed the self-diffusion coefficients in methane on the same grid as the EOS described in Sec.~\ref{sec:methane} as well as in the 2:1:4 ternary mixture along the three 
planetary profiles described in Sec.~\ref{sec:Uprofiles}. Data for the self-diffusion coefficients in water and ammonia were already available in the literature 
\citep{French2010,Bethkenhagen2013}, which we then complemented with additional calculations as necessary. The results for the four considered species are shown in 
Fig.~\ref{fig:diff}. Additionally, the diffusion coefficients of each species in the real ternary mixture as well as radius, density, temperature, and pressure along the 
three profiles are given in Tab.~\ref{tab:diff}. In general, the diffusion coefficients in the ternary mixture agree within 20\% with those in 
the pure compounds, which is a satisfactory result, given that the numerical uncertainty is usually 5-10\%. The most significant deviations are found for conditions where 
the underlying phases in the pure compounds and mixtures differ significantly from each other. This occurs, for example, in regions close to rotationally-disordered methane 
or in the superionic phase of water, whereas the ternary mixture is characterized as an ordinary fluid. This effect can be seen for the diffusion coefficients of carbon, 
nitrogen, and oxygen in Fig.~\ref{fig:diff}. In those regions, the diffusion coefficients of these heavy particles vanish (for illustration purposes artificially set to 
10$^{-5}$ cm/s$^2$), while that in the ternary mixture retain values typical for a fluid.

\section{An Icy Uranus Model}\label{sec:IcyUranusModel}

Wide-range equations of state for real mixtures of icy materials are generally not available. In Sec.~\ref{sec:LMA}, we have quantified the uncertainty 
of applying the LMA to the single component EOS for selected icy mixtures. The error was found 
to be of the order of a few percent in density and a few kJ/g in internal energy or smaller, in particular in regions where 
all single components are in the fluid phase and off regions of phase changes or signs of demixing. 
Although these requirements do not entirely hold along the cool Uranus and Neptune adiabats \citep{French2009,Hirai2009,Chau2011}, we are interested in the effect of 
applying our {\it ab initio} EOS of the icy mixture to the inner mantle of Uranus in comparison to former work that relied only on water as a representative for all ices 
\citep{Redmer2011,Helled2011,Nettelmann2013}. Thus, we here linearly mix the EOS of methane, ammonia, water, hydrogen, and helium and compute a new Uranus model where heavy 
elements in the deep interior are represented by a solar mixture of H:C:N:O (Anders and Grevesse 2009) in form of methane, ammonia, and water, yielding a metallicity ratio
$Z_{\rm CH_4}:Z_{\rm NH_3}:Z_{\rm H_2O} = 0.31:0.08:0.61$.  Otherwise, our interior structure modeling procedure follows exactly that of Nettelmann et al (2013), i.e.,~we 
assume three layers where the a priori unknown heavy element mass fractions $Z_1$ and $Z_2$, in the two adiabatic H/He/ice envelopes are used to adjust the gravitational 
harmonics $J_2$ and $J_4$, while the rock core mass is used to account for total mass conservation. We use the rotation rate from the Voyager mission. Of course, the real 
interior structure of Uranus may be far more complex than our model. 

\begin{figure}[htb]
\centering
  \vspace{-4pt}
\includegraphics[width=0.82\columnwidth]{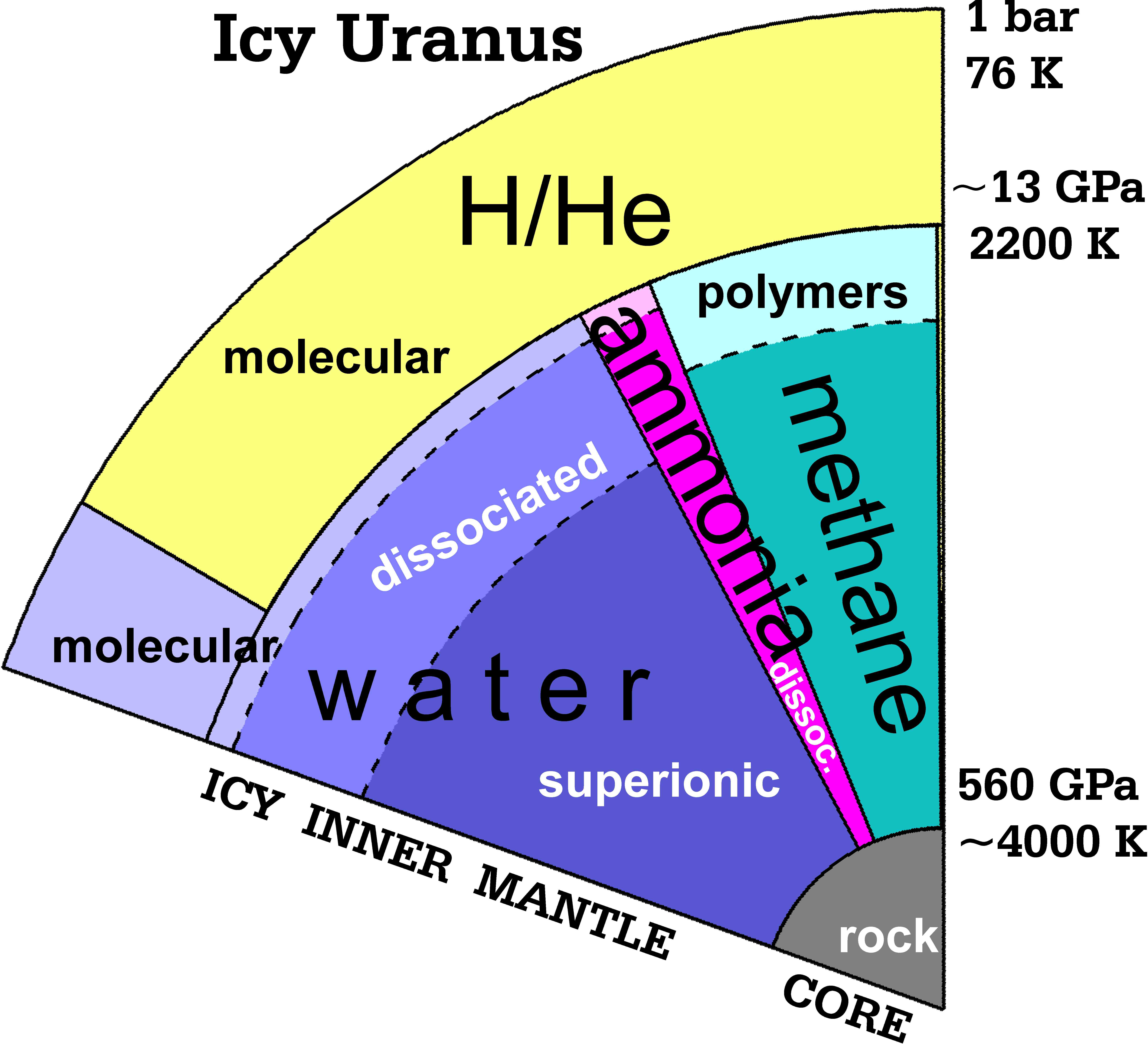}
\caption{\label{fig:UranusTorte}
Icy Uranus structure model with three homogeneous layers (\emph{solid} azimuthal lines). The radial direction scales linearly with the planet's internal radial distance from 
the center, while the angle scales with mass abundance of the single components. \emph{Dashed} lines indicate phase boundaries of the single components according to the 
phase diagrams of water \citep{Redmer2011}, ammonia \citep{Bethkenhagen2013}, and methane \citep{Hirai2009}, except the \emph{light red} color which indicates partial 
dissociation of ammonia into $N_2$ and $H_2$.} 
\end{figure}

Our resulting icy Uranus model shows four distinct features: high-$Z_2$ values of 0.98-0.999, i.e.,~an almost purely icy deep interior, low central temperatures of 
$T_{\rm core}\sim 4000$ K, high ice:rock ratio of $\sim 19$, and a narrow range of possible transition pressures $P_{1-2}=10$--15 GPa between the outer H/He-rich and the 
inner ice-rich envelope. Figure \ref{fig:UranusTorte} illustrates the icy Uranus model for $P_{1-2}=13$ GPa. Our icy Uranus model turns out to be similar to that of 
\citet{PodRey1987}, who applied linearly mixed EOSs for the ices H$_2$O, CH$_4$, NH$_3$ and H$_2$S based on Thomas-Fermi-Dirac theory for dense matter and an interpolation 
to ideal gas region at the low pressures. In addition, they took into account the influence of condensation on the temperature profile in the outer envelope. They assumed an 
ice shell atop a rock core and found a high ice:rock ratio of 16.6 necessary to explain the measured $J_2$ and $J_4$ values of that time.

For comparison, planet models where HCNO-bearing molecules are represented by a water EOS  typically predict lower inner envelope ice mass fractions of $Z_2\sim 0.9$ for 
Uranus. This result appears to be independent of whether a three-layer model approach \citep{Hubbard1995}, a smooth density distribution \citep{Helled2011}, or random 
interior structure search for acceptable density distributions \citep{Podolak2000} was used. The higher $Z_2$ value of our icy Uranus model is a direct consequence of the 
lower mean molecular weight of methane and ammonia compared to that of water, reducing the need for adding H/He. The then reduced number of light atoms (helium, hydrogen) 
leads to an increased specific heat, which tends to reduce the temperature along adiabats. As a result, we obtain $T_{\rm core}$ values around 3800~K only, with an 
uncertainty of about $\pm$ 500~K resulting partially from the method of computing isentropic $P$--$T$ profiles. The temperature along our icy Uranus adiabat might thus be a 
lower bound to that of a real mixture adiabat.

It can be argued that this icy Uranus model may \citep{HubbardMacfarlane1980,PodRey1984} or may not \citep{PodRey1987} be in conflict with predictions from formation theory 
\citep{Pollack1996}, or be too warm \citep{Hubbard1995} or too cold \citep{Nettelmann2016} to explain the observed low luminosity. However, the solid conclusion that can be 
drawn from this extreme model is: \emph{if} the ices adopt a solar mixing ratio and \emph{if} deep internal temperatures are signifcantly higher than $\sim 4000$~K, the inner 
mantle \emph{must} contain also heavier, rock-forming elements mixed with the ices, and \emph{can} contain more H/He. Therefore, it will be important to investigate the 
mixing behavior of planetary ices with rocks and H/He in the future.

\section{Conclusions}

Overall, we find the linear mixing approximation to perform remarkably well for the molecular compounds methane, ammonia, and water, under the thermodynamic conditions 
predicted in the mantle of Uranus. The maximum deviation between the three computed real 1:1 binary mixtures and the linear mixing model amounts to 4\% in density and -4~kJ/g
in specific internal energy. The latter corresponds to 4\% deviation in the planetary radius coordinate or a 200~K shift in temperature. Note, that this deviation is smaller 
than the uncertainty in the temperature profile related to the interior models. Even smaller discrepancies of at most 2.1\% in density and -2.9~kJ/g in internal energy were 
observed for the 2:1:4 ternary mixture along three representative Uranus profiles. These particular deviations have a characteristic sign: the LMA overestimates the density, 
while the internal energy is underestimated. If the same thermodynamic phase is present in both the real mixture and the pure compounds, the linear mixing approximation 
performs even better than stated above. Hence, future work will be directed toward the construction of reliable thermodynamic potentials for the pure compounds water, ammonia,
and methane, since it does not seem necessary to construct many EOS for different compositions. Our new methane EOS together with the 
water~\citep{French2009, French2015, French2016} and ammonia~\citep{Bethkenhagen2013} data tables that were extended here, will provide an excellent starting point for that. 

Moreover, the diffusion coefficients for the individual species in the mixture were found to agree within 20\% with the values observed in the pure compounds as long as the 
same state of matter is present. This implies that accurate knowledge of the phase diagrams of pure components as well as their mixtures is still required to understand the 
planetary interiors even though the linear mixing approximation works well. For example, the survival of a superionic phase in the presence of methane is still an open 
question. Also the potential demixing of methane into carbon and hydrogen needs further investigation. The formation of polymers observed in our methane simulations might 
already hint into that direction. A deeper understanding of superionicity or demixing phenomena in planetary H-C-N-O mixtures will be beneficial for the future development 
of more advanced planetary models with a more complex interior structure. 

The ice-rich Uranus model introduced here illustrates the lower temperature bound of possible interior structure models. However, the model does not recover the correct age 
of Uranus and needs to be improved further. Thermal-boundary-layer (TBL) models~\citep{Nettelmann2016}, such as the hottest model considered here, might be a promising 
alternative. It will be insightful to provide these TBL models with well-founded input quantities in the future, especially transport properties, such as the viscosity and 
the electrical and thermal conductivity. These properties are also of great interest for dynamo simulations~\citep{Wicht2010}. For example, it would be desirable to have a 
complete set of transport and thermodynamic properties along the discussed planetary profiles, similar as that for the hydrogen-helium mixtures along the Jupiter 
adiabat~\citep{French2012}.

\section*{Acknowledgment}

We thank R. Helled, M. Podolak, C. Kellermann, and M. Sch\"ottler for insightful discussions. 

MB, NN, MF, and RR gratefully acknowledge support from the Deutsche Forschungsgemeinschaft within the SFB 652, the SPP 1488, and the FOR 2440. Computing power was provided 
by the ITMZ of the University of Rostock as well as by the North-German Supercomputing Alliance (HLRN). ERM, CT, JDK, and LAC gratefully acknowledge support from the 
Advanced Simulation and Computing Program (ASC), science campaigns 1 and 4, and LANL which is operated by LANS, LLC for the NNSA of the U.S. DOE under Contract 
No. DE-AC52-06NA25396. SH and MB acknowledge support by the U.S. Department of Energy at the Lawrence Livermore National Laboratory under Contract No. DE-AC52-07NA27344 and 
the LDRD grant 16-ERD-011.

\bibliographystyle{apj}
\bibliography{linMix}

\end{document}